\documentclass[a4paper,11pt]{article}
\pdfoutput=1 % if your are submitting a pdflatex (i.e. if you have
\usepackage{jinstpub} % for details on the use of the package, please
\usepackage{comment}

\usepackage{subcaption}
\usepackage{bm}
\usepackage{listings}
\usepackage{adjustbox}

\title{Investigations on a Fuzzy Process: Effect of Diffusion on Calibration and Particle Identification in Liquid Argon Time Projection Chambers}
\begin{document}

\author[b]{A.~Lister}
\author[a]{and M.~Stancari}

% Institutions in alphabetical order
\affiliation[a]{Fermi National Accelerator Laboratory\\Kirk Rd \& Pine St W, Batavia, IL, USA}
\affiliation[b]{University of Wisconsin - Madison\\ Chamberlin Hall, 1150 University Avenue, Madison, WI, USA}

\abstract{Ionization electron diffusion in Liquid Argon Time Projection Chambers (LArTPCs) has typically been considered at the detector design stage, but little attention has been given to its effects on calibration and particle identification. We use a GEANT4-based simulation to study how diffusion impacts these techniques, and give consideration to how this effect is simulated. We find that diffusion can cause a drift-dependent bias to both the median and Most Probable Value (MPV) of d$Q$/d$x$ distributions. The bias is estimated to be $\sim2.5$\% (median) and $\sim5.0$\% (MPV) for typical maximum drift times in currently running LArTPCs before adding detector specific considerations such as electric field non-uniformities. This indicates that these metrics should not be used for calibration without care, contrary to the conventional wisdom. The impact of diffusion on the ability of LArTPCs to separate muons and protons is small, and not expected to pose any problems in future detectors. Diffusion may however be a significant source of systematic uncertainty when separating particles of more similar masses (muons and pions, kaons and protons). Separation of such populations may be improved by implementation of a drift-time dependent particle identification.}


%\arxivnumber{1234.56789} % only if you have one

\emailAdd{adam.lister@wisc.edu}
\date{October 2021}
\maketitle

%%%%%%%%%%% SECTIONS %%%%%%%%%%%%%

\section{Motivation}

Clouds of free electrons in a medium will spread out isotropically over time due to diffusion. In a Liquid Argon Time Projection Chamber (LArTPC) ionization electrons drift under the influence of an electric field ($E$-field), meaning that diffusion becomes non-isotropic. It is often parametrized in terms of components longitudinal ($D_L$) and transverse ($D_T$) to the $E$-field. Measurements of diffusion at the $E$-fields relevant for LArTPCs, where the parameters should be relatively constant as a function of $E$-field, have resulted in $D_L$ $\sim 5 \pm 1.5$ $\mathrm{cm}^2/\mathrm{s}$ \cite{MicroBooNE:2021icu, Lister:2019kmq, Cennini:1994ha, Li:2015rqa}, but no measurements have been made of the transverse component at these $E$-fields. Instead, the value of $D_T$ used in current simulations is inferred from measured values of $D_L$, however there is tension in the currently measured values of $D_L$ \cite{MicroBooNE:2021icu}. 

Typically diffusion has been considered during detector design where the signal-to-noise ratio is estimated using diffusion combined with the wire spacing, $E$-field, argon purity, and noise level. So far as we are aware, there are no published investigations into the effects of diffusion on energy reconstruction, calibration, and particle identification (PID). In particular, the transverse component of diffusion, which can lead electrons to be spread across multiple readout channels, may lead to an averaging effect with the potential to modify the shape of the observed energy deposition per unit length (d$E$/d$x$). Because the d$E$/d$x$ changes rapidly close to the Bragg Peak, it might be expected that such an effect may distort the shape of the d$E$/d$x$ distribution in this region more than elsewhere, impacting the particle identification capabilities of LArTPCs. Further, because diffusion spreads electron clouds in a stochastic way in three dimensions, with a magnitude that is drift-time dependent, it cannot be effectively calibrated in the same way that effects such as electron attenuation and recombination can be calibrated. Some proposed future detector designs have large drift distances, up to 6.5 m \cite{Paulucci:2021sqn}. For long drift-time detectors, quantifying how diffusion degrades the detector performance near the cathode is of utmost importance. Further, it is important that we consider whether the impact of diffusion on energy calibrations must be accounted for to meet the strict energy scale uncertainty requirement (~1-2\%) for the Deep Underground Neutrino Experiment (DUNE) to attain its physics goals \cite{DUNE:2020ypp}. In this article, we introduce a simulation of diffusion into GEANT4 \cite{GEANT4:2002zbu} to study its effect.

% removed discussion of 12 m drift since it's no longer being suggested  \cite{DUNE:2020lwj})

\section{Simulation Details}

The simulation used throughout this work is based on GEANT4, with custom routines to apply the longitudinal and transverse components of diffusion. The detector is simulated as a single block of liquid argon. The $x$-direction is digitized using the simulated width of the detector, $E$-field (500 V/cm), and clock frequency (2 MHz), and the $z$-direction is then split into detection regions (``wires'', 3 mm) to approximate a collection plane. For this study neither induction planes nor $E$-field distortions are considered. For all simulations reported here, we use the \lstinline{QGSP_BERT} physics list. To ensure multiple sets of ionization electrons land in each detection region, we limit the maximum allowed size of each step GEANT4 takes to 0.3 mm.

\begin{figure}[h!]
\centering
\includegraphics[width=0.49\textwidth]{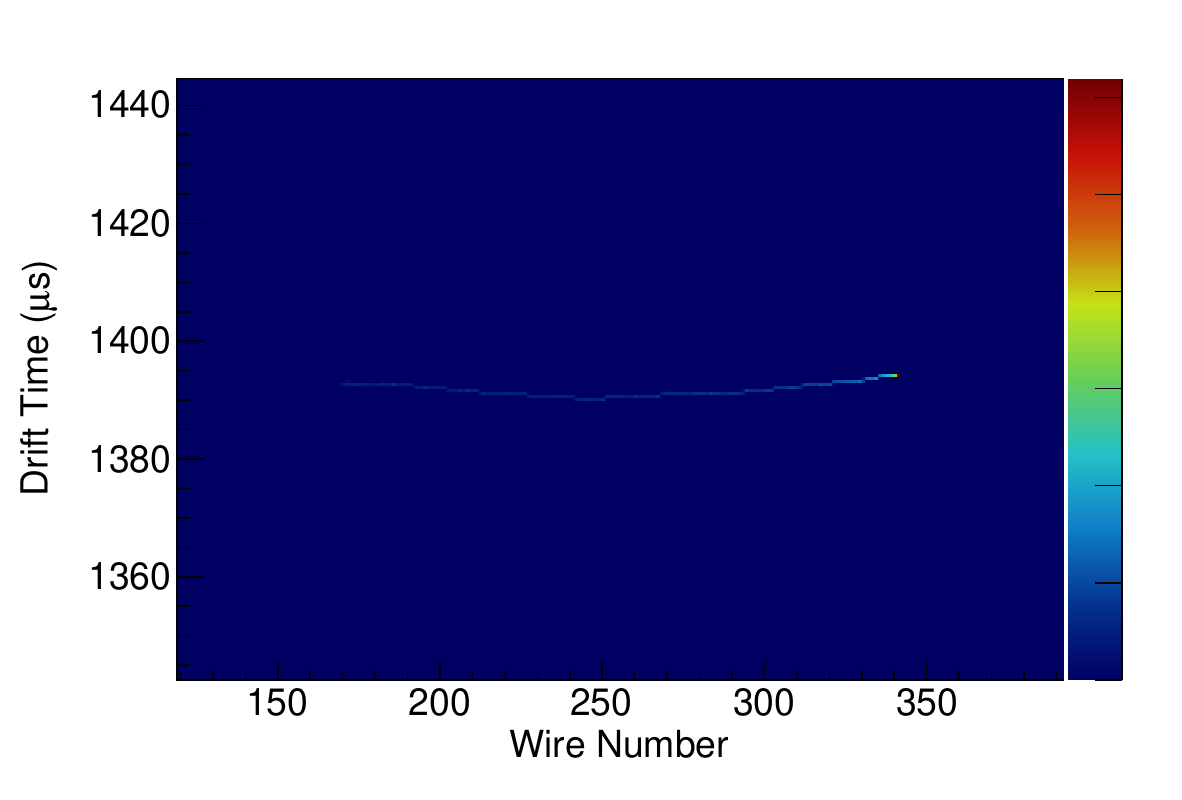}
\includegraphics[width=0.49\textwidth]{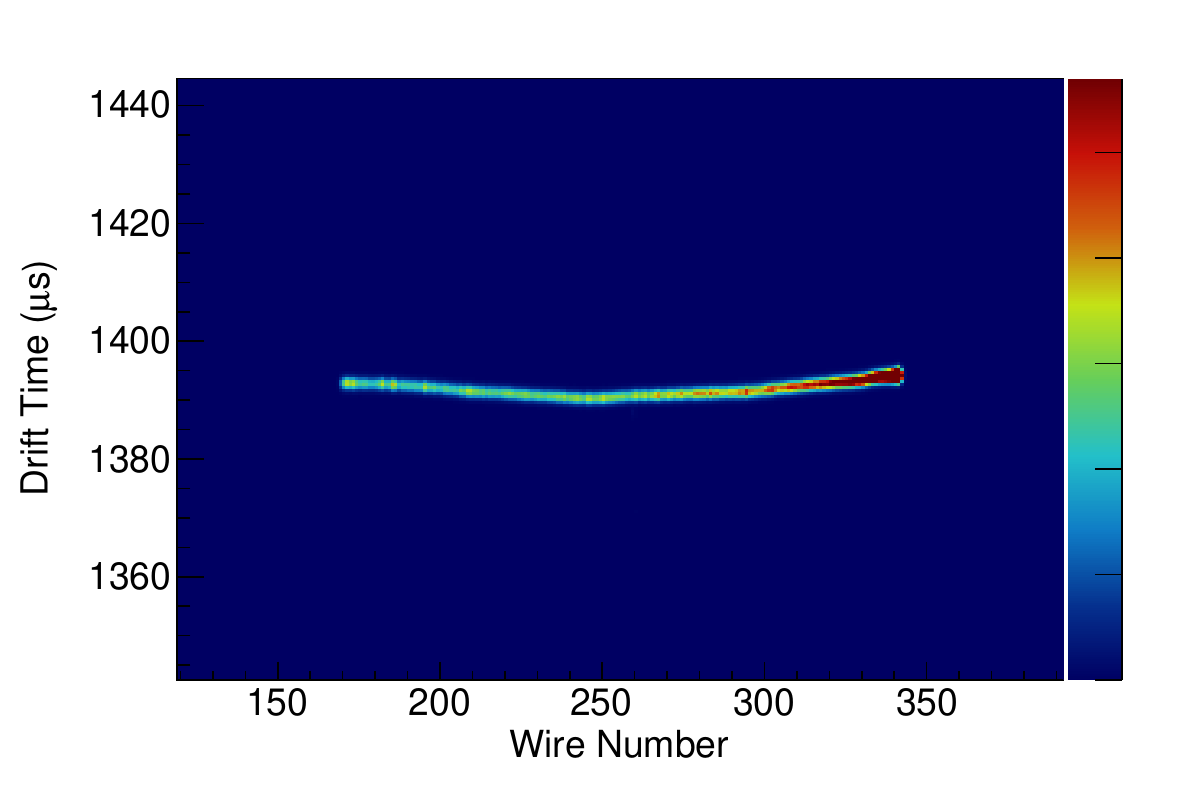}
\caption{Event displays of a simulated 0.3 GeV proton before (left) and after (right) applying transverse and longitudinal diffusion.}
\label{fig:evds}
\end{figure}

\begin{figure}[h!]
\centering
\includegraphics[width=0.6\textwidth]{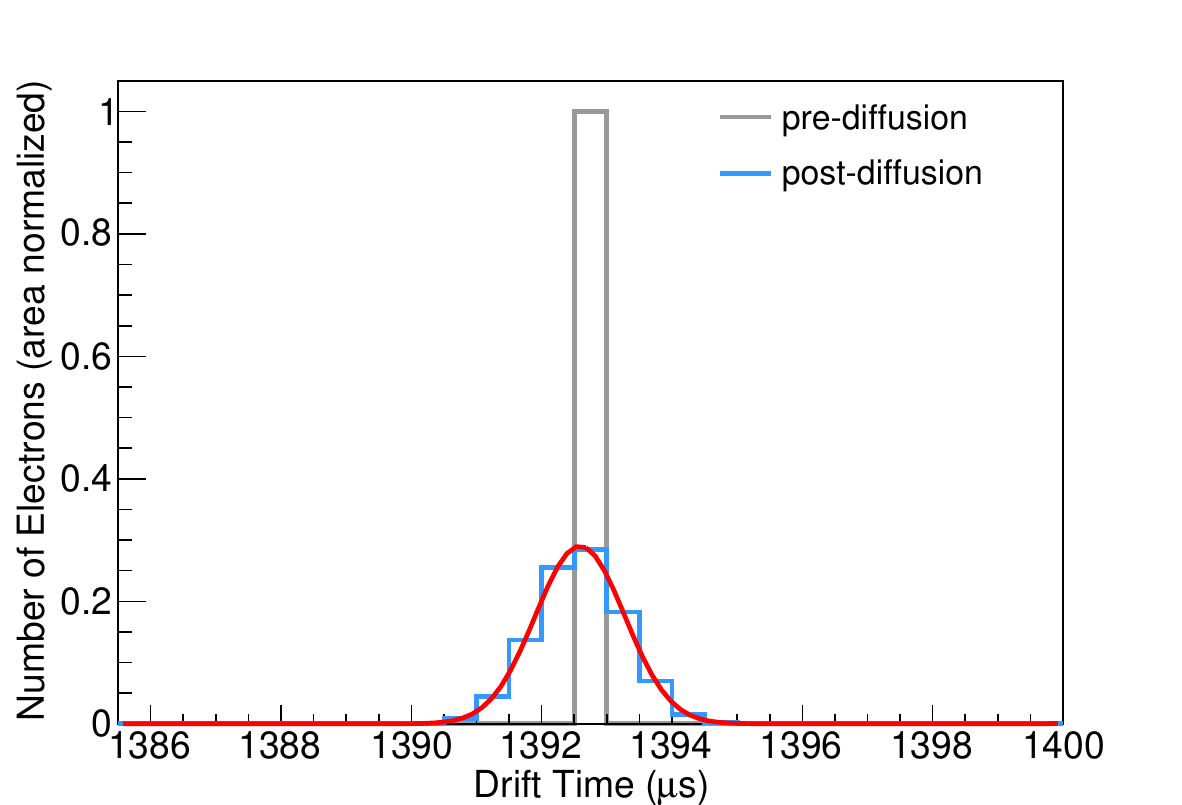}
\caption{Distribution of electrons on a single wire before and after applying diffusion. After applying diffusion the signal becomes Gaussian-like as evidenced by the Gaussian fit, shown for comparison.}
\label{fig:waveforms}
\end{figure}

For each event, a single particle is simulated in the detector, depositing energy in each step. We apply recombination via the modified Box model \cite{ArgoNeuT:2013kpa} using the d$E$/d$x$ in each step,
\begin{equation}
    \frac{\mathrm{d}Q}{\mathrm{d}x} = \frac{\ln[\beta \frac{\mathrm{d}E}{\mathrm{d}x} + \alpha]}{\beta W_{\mathrm{ion}}},
\end{equation}
\noindent where $\alpha=0.93$, $\beta=0.3$ cm/MeV, and $W_{\mathrm{ion}}$ = 23.6 eV is the mean ionization energy of liquid argon. The calculated d$Q$/d$x$ is used to generate a number of electrons produced at each step.  

The electrons in the step are assigned a drift time, $t_d$, based on the true distance of the ionization electrons from $x=0$ and the simulated $E$-field. The electrons may then be subdivided into a number of packets to increase the speed of simulation, and each electron packet is assigned a new position based on its drift time and the chosen values of $D_L$ and $D_T$. First, the magnitude of the shift due to the different diffusion components is calculated by randomly sampling Gaussian distributions centered at 0 with,

\begin{equation}
    \sigma_L (\mu s) = \sqrt{\frac{2D_L \cdot t_d}{v_d^2}},
\end{equation}
\noindent and
\begin{equation}
    \sigma_T (mm) = \sqrt{2D_T \cdot t_d},
\end{equation}

\noindent where $v_d$ is the drift velocity \cite{Li:2015rqa}. With the magnitudes $m_L \sim G(0, \sigma_L)$ and $m_T \sim G(0, \sigma_T)$, we choose a random number between 0 and $\pi$ for the new position in the $y$, $z$ direction, $d=\mathrm{Rand(0, \pi)}$. New positions are then assigned,

\begin{align}
    t_d^{new} &= t_d + m_L, \\
    y^{new}   &= y + m_T\cos{d}, \mathrm{ and} \\
    z^{new}   &= z + m_T\sin{d}.
\end{align}

\noindent A closure test is performed by ensuring that we are able to extract the correct value of $D_L$ from the simulation. As a demonstration of the simulation of diffusion, example event displays before and after diffusion are shown in figure \ref{fig:evds}, and an example of the distribution of electrons on a wire before and after applying diffusion is shown in figure \ref{fig:waveforms}. 

\section{Reconstruction of d\textit{Q}/d\textit{x} and d\textit{E}/d\textit{x}}

To remove effects of poor reconstruction from this study we do not apply any reconstruction algorithms, and instead the path of the true initial particle is saved, producing a track which can be considered to be \textit{perfectly reconstructed}. Secondary particles are not saved; this choice removes any showering particles from consideration, but also has the effect of removing the end of tracks where GEANT4 considers a new particle to have been produced after a scatter. 

The number of electrons collected on a wire is calculated by integrating the number of electron packets in a time window around the true trajectory, $N$, and multiplying by the packet size, $P$, d$Q = NP$. The distance between the current energy deposition and the energy deposition on the preceding wire is calculated using the true trajectory, and used to construct the d$Q$/d$x$. The d$E$/d$x$ is then calculated, again using the modified Box model. The distance to the true end point of the trajectory of the particle is also calculated and stored.

\section{Results and Discussion}

\subsection{Samples Used}
\label{subsec:samples}

We use two samples to estimate the effects of diffusion on calibration and particle identification
\begin{itemize}
\item \textbf{Idealized Sample}: Simulated 1 GeV muons and 0.3 GeV protons, along the $z$-axis, that are only allowed to undergo ionization, with all other interactions turned off. This sample is used so that we're able to isolate the effects of diffusion from the other effects which may complicate analysis such as delta-ray production and high track angles from particle scatters, each of which can modify the reconstructed  $\mathrm{d}Q/\mathrm{d}x$. 
\item \textbf{Physics-on Sample}: Simulated 1 GeV muons, 0.3 GeV protons and kaons, and 0.1 GeV pions, along the $z$-axis, which have all physics in the physics list turned on. This sample is used for studying particle identification.
\end{itemize}

\noindent The energy of the muons is relatively unimportant given they do not interact hadronically. The energy of the protons, kaons, and pions were chosen to maximise the length of the track while ensuring most of the particles do not re-interact. Both of these samples use $D_L = 6.5$ cm$^2$/s and $D_T = 13$ cm$^2$/s and have a wire pitch of 3 mm. This configuration was chosen to be conservative, such that any lower values of the diffusion constants, or larger wire spacing should result in a smaller effect than what is presented in this study.  

\begin{figure}[t]
\centering
\includegraphics[width=0.6\textwidth]{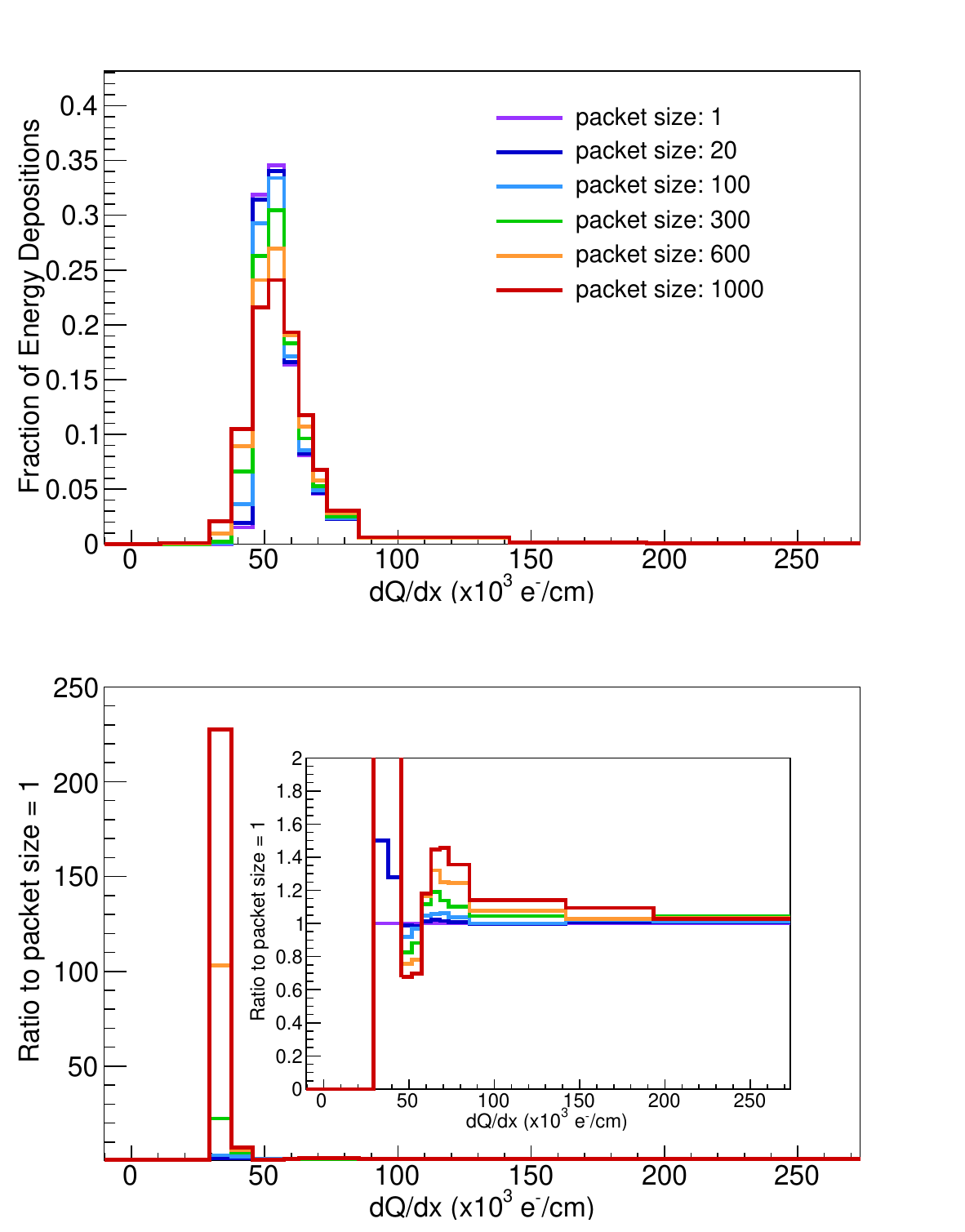}
\caption{d$Q$/d$x$ distributions using different packet sizes. The shape becomes increasingly distorted as the packet size increases. The bottom plot shows the ratio to the distribution using a packet size of 1, and the inset plot shows the same ratio but limited to the region [0,2].}
\label{fig:electron_packet_size}
\end{figure}

Because there are $\mathcal{O}$(1000)s ionization electrons liberated per mm by charged particles travelling through liquid argon, simulating diffusion for each electron is time consuming. In some current simulations of diffusion, electrons are therefore grouped together into packets of configurable size, and all electrons in a packet are assigned the same new position after diffusion. We have investigated the effects of packet size on the d$E$/d$x$ distribution in figure \ref{fig:electron_packet_size} and find that it can have a significant impact on the shape of the distribution. We advocate for simulations using this approach to keep packet sizes to a minimum. Increasing the packet size to be larger than $\sim 20$ results in changes to bin contents of order 5\% or larger. This effect is caused by relatively rare large values sampled from the Gaussian distribution being applied to each of the electrons in the packet rather than just a single electron. For this reason, we use an electron packet size of 1, meaning we individually apply diffusion to every electron.

\subsection{Impact of Diffusion on Minimum Charge Efficiency}

\begin{figure}[ht]
\centering
\includegraphics[width=1.0\textwidth]{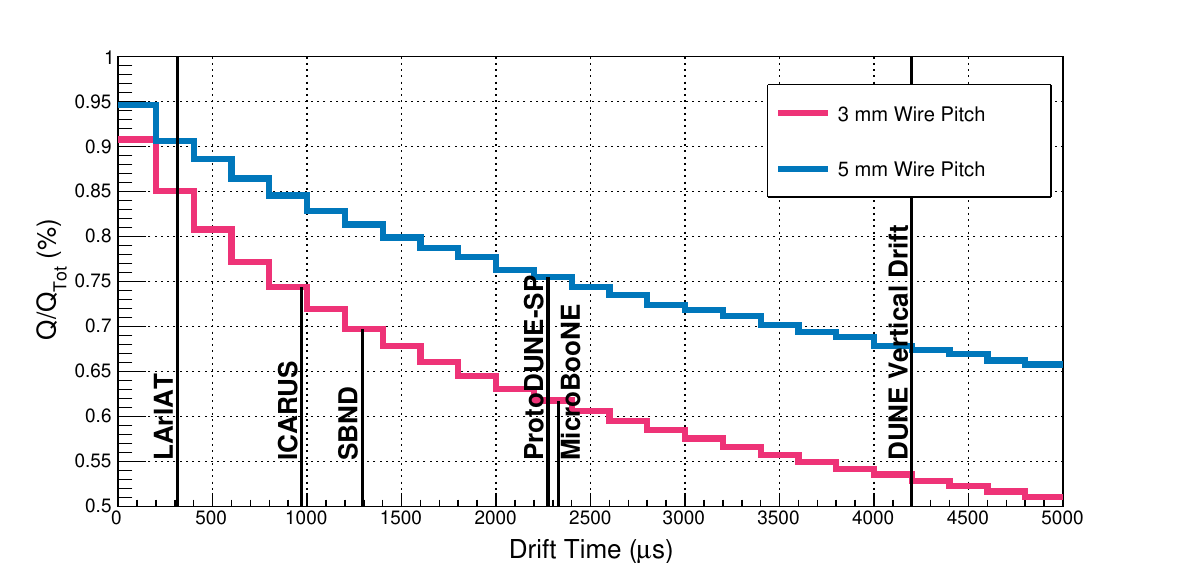}
\caption{Minimum charge efficiency using idealized muons in a simulated detector with 3 mm wire spacing and 5 mm wire spacing. Drift times for each experiment are estimated using their maximum drift distances, $E$-fields, and temperatures \cite{LArIAT:2019kzd,ICARUS:2004wqc,SBND:2020scp,DUNE:2020cqd,MicroBooNE:2019lta} in conjunction with the parametrization of the electron mobility introduced in reference \cite{Li:2015rqa}. Where operating temperatures aren't known (SBND and DUNE vertical drift), 89 K is assumed. Note that LArIAT has different values for wire pitch, and slightly different $E$-fields and operating temperatures in each run \cite{LArIAT:2019kzd}.}
\label{fig:chargeresolution}
\end{figure}

Using the idealized sample outlined in \S\ref{subsec:samples}, we are able to look at the minimum charge efficiency, $Q/Q_{Tot}$, which describes the fraction of electrons produced at the cathode which land in the detection region that would be expected without diffusion. We emphasise here that diffusion only moves ionization electrons around, meaning the total energy deposited is conserved, only the apparent location of the depositions changes. Figure \ref{fig:chargeresolution} shows that with the assumption of $D_T$ = 13 cm$^2$/s the minimum charge efficiency varies wildly between running experiments, with LArIAT having the highest minimum charge efficiency, greater than $90$\% in some configurations, and MicroBooNE having the lowest minimum charge efficiency ($\sim60$\%). We also include the current configuration for the DUNE vertical drift module (6.5 m, $E$-field of 500 V/cm), though this detector will not use sense wires. Aside from LArIAT, which benefits from its short drift distance, the maximum expected minimum charge efficiency for any detector is $\sim 75\%$. Clearly, 25\% of the expected charge being deposited on a channel other than the expected one for large drift times is a strong motivator for this study.

\subsection{Impact of Diffusion on dQ/dx Distributions}
\label{subsec:dedximpact}

\begin{figure}[h!]
\centering
\includegraphics[width=0.49\textwidth]{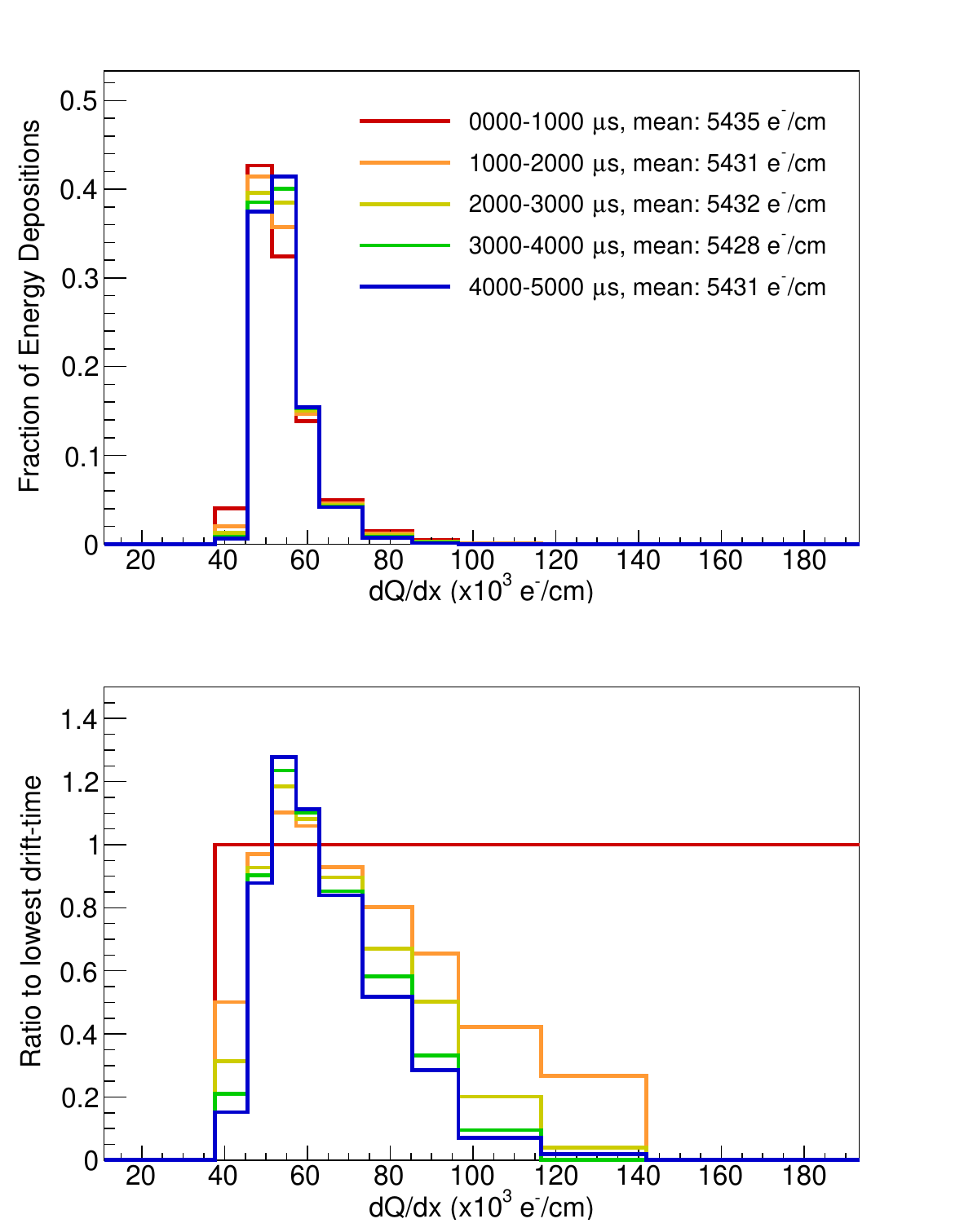}
\includegraphics[width=0.49\textwidth]{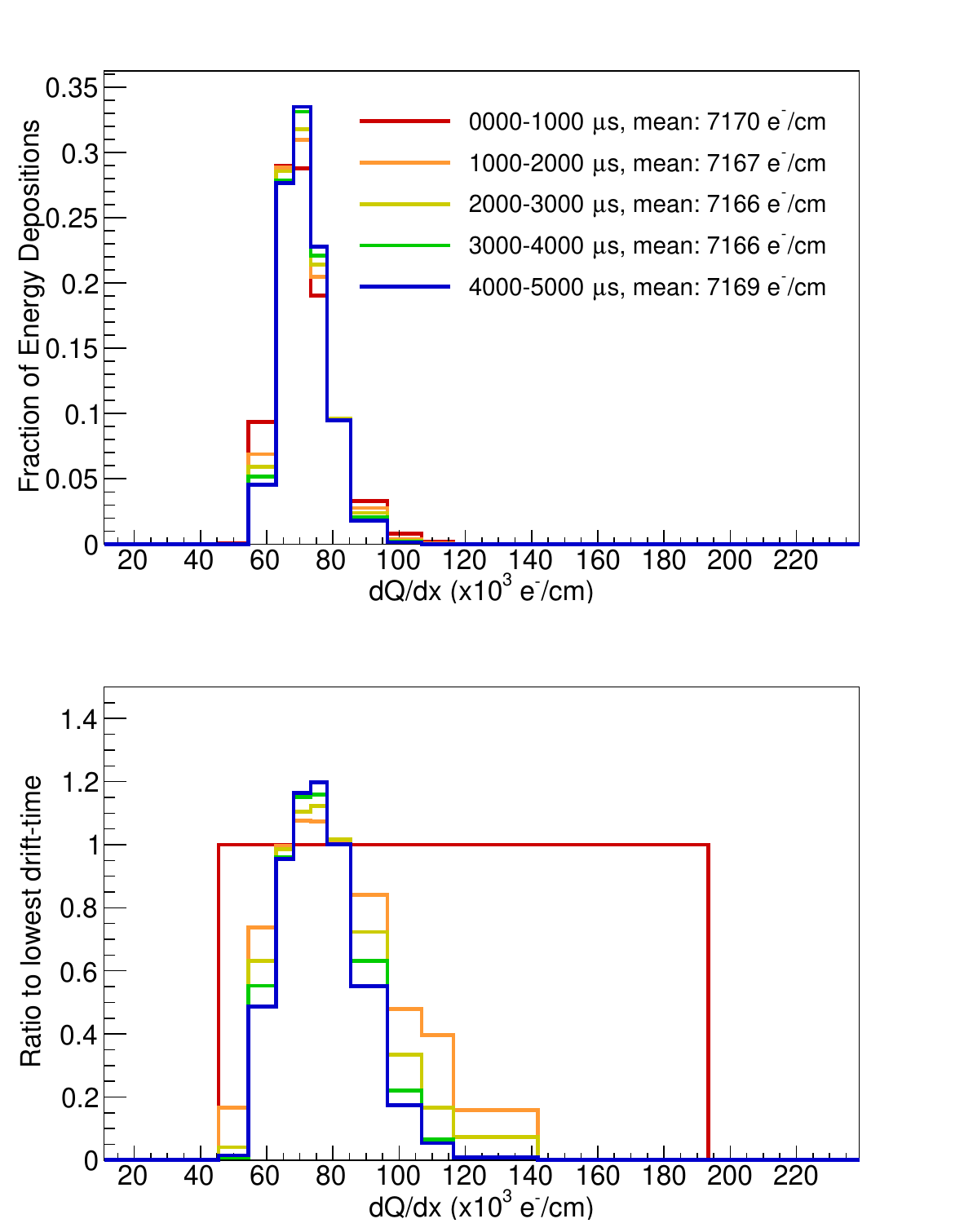}
\caption{Area normalized distributions of d$Q$/d$x$ for the MIP-region (a 1 m region 1 m from the end of the track) for muons (left) and protons (right) as a function of drift time using perfectly forward going particles with only ionization simulated. The bottom plots show ratio of each drift bin to the lowest drift bin. The distributions become distorted as a function of drift time, with the MPV moving to the right for greater drift times, however the mean of the distributions remains relatively constant.}
\label{fig:1dim}
\end{figure}

Distributions of d$Q$/d$x$ for idealized muons and protons at different drift distances can be found in figure \ref{fig:1dim}. Diffusion acts to modify the shape of the distributions as a function of drift time, however the means of the distribution do not meaningfully change. Diffusion also averages out the stochastic nature of energy deposition, bringing the distribution closer to the mean. This can be understood by considering a toy experiment with only two wires. A Landau-distributed random number is drawn for each wire to simulate a charge deposition, and then diffusion is simulated, resulting in some fraction of charge, $A$, from the first wire landing on the second wire, and vice-versa. The charge collected on the first wire after diffusion, $C_{w1}^{post}$ can then be considered as a linear combination of the charge deposited on each wire before diffusion ($C_{wN}^{pre}$), $C_{w1}^{post} = A \times C_{w1}^{pre} + (1-A) \times C_{w2}^{pre}$. A linear combination of numbers drawn from the same random distribution does not follow the input distribution, but actually becomes more tightly concentrated around the mean of the distribution\footnote{With far more than two samples, the distribution would asymptotically approach a Gaussian with a mean of the original Landau distributions according to the central limit theorem.}. In general $\mathrm{d}E/\mathrm{d}x$ distributions are known to change as a function of $\mathrm{d}x$ - an effect which is noted around figure 34.8 of reference \cite{Zyla:2020zbs}. This effect is due to the larger chance of a hard scatter with an increased d$x$, pulling the measured d$E$/d$x$ to larger values. This process is similar in effect, but meaningfully different from what we observe, in that what we see is not due to the width of the detection region but an averaging effect due to the physical mixing of electrons between detection regions.

\subsection{Impact of Diffusion on Calibration}

\begin{figure}[ht]
\centering\includegraphics[width=1.0\textwidth]{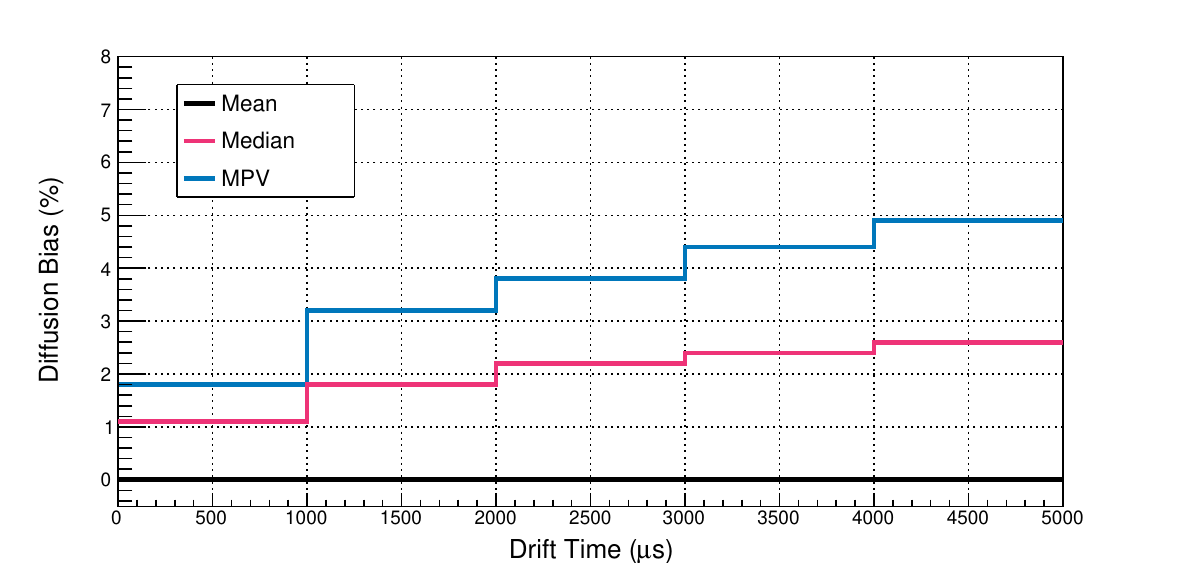}
\caption{Bias of mean, median, and MPV d$Q$/d$x$ values after applying diffusion when compared to the true mean, median and MPV pre-diffusion. The idealized sample of 1 GeV forward-going muons outlined in \S\ref{subsec:samples} was used to estimate the bias.}
\label{fig:biasplot}
\end{figure}

The d$Q$/d$x$ distributions changing as a function of drift time has implications for how LArTPCs are calibrated. Current calibration techniques for LArTPCs primarily use cosmic ray muons as calibration sources. The median and most probable value (MPV) of d$Q$/d$x$ distributions are used rather than the mean to be more robust against the effect of delta rays which might overlay the tracks \cite{MicroBooNE:2019efx,DUNE:2020cqd}, and other reconstruction pathologies. Because diffusion acts to move the distributions closer to the mean, and because the distribution is non-Gaussian with a tail to high d$Q$/d$x$ values, the median and MPV both move to higher d$Q$/d$x$ values as a function of drift time. Using an idealized sample of muons, we can see the effects of the bias in table \ref{table:meanmeddedxvals} (shown graphically in figure \ref{fig:biasplot}). The mean values of d$Q$/d$x$ after diffusion do not change when compared to the same distributions pre-diffusion, however the median and MPV values change significantly, with a bias of $\sim5$\% in MPV for the those depositions with the longest drift times considered. We have also studied muons generated at thirty (sixty) degrees to the plane and found that the bias due to diffusion remained approximately constant. There is an additional bias of 1.9\% (3.3\%) incurred from the increased width of material the muon travels through for each detection region for these angular samples. We note here that the bias we report is dependent on the values of the diffusion and recombination parameters chosen for the simulation and should be used only as an example of the effect.

\begin{table}
\begin{center}
\begin{adjustbox}{width={\textwidth},totalheight={\textheight},keepaspectratio}%
\begin{tabular}{c||c|c|c||c|c|c||c|c|c}
      Drift Time ($\mu$s) & \multicolumn{3}{|c||}{Mean d$Q$/d$x$ (e$^-$/ cm)} & \multicolumn{3}{|c||}{Median d$Q$/d$x$ (e$^-$/cm)} & \multicolumn{3}{|c}{MPV d$Q$/d$x$ (e$^-$/cm)}\\
     \hline
     & no diff & diff & \% bias & no diff & diff & \% bias & no diff & diff & \% bias\\
    \hline
    \hline
    0-1000     & 54358.3 & 54359.1 & 0.0 & 51556.7 & 52108.6 & 1.1 & 49003.5 & 49909.2 & 1.8\\
    1000-2000  & 54313.9 & 54313.9 & 0.0 & 51492.1 & 52426.7 & 1.8 & 48921.9 & 50491.1 & 3.2\\
    2000-3000  & 54325.4 & 54326.2 & 0.0 & 51521.1 & 52659.6 & 2.2 & 48934.4 & 50797.2 & 3.8\\
    3000-4000  & 54287.5 & 54287.4 & 0.0 & 51535.3 & 52769.0 & 2.4 & 48962.7 & 51093.3 & 4.4\\
    4000-5000  & 54315.9 & 54316.2 & 0.0 & 51531.9 & 52870.8 & 2.6 & 48826.3 & 51202.1 & 4.9\\
\end{tabular}
\end{adjustbox}
\end{center}
\caption{Mean, median, and MPV d$Q$/d$x$ values before and after applying diffusion for different drift time ranges. These were made using an idealized sample of 1 GeV forward-going muons outlined in \S\ref{subsec:samples}.}
\label{table:meanmeddedxvals}
\end{table}

Current response-flattening ($\mathrm{d}Q/\mathrm{d}x$) calibration techniques could be impacted by the effect discussed above. Calibrations in the Y-Z plane may become biased if the $x$-direction coverage is not approximately equal in each bin, but the larger, more direct impact on such a calibration strategy is on the $x$-direction response-flattening calibration which is almost certainly biased by this effect. This effect also likely impacts current energy-scale (d$E$/d$x$) calibrations, which use the MPV value of d$Q$/d$x$. Given that the mean of the distribution appears to be an unbiased estimator across the volume, one could consider using the mean in calibration efforts. The presence of delta rays, which would modify the shape of the distribution, should bias the mean by the same amount across the drift time since delta rays should be produced equally across the volume (see appendix \ref{app:calibrationwithphysicson}). 

Either way we choose to calibrate, we introduce bias. Current techniques using the median and MPV of the distribution are biased in a drift-dependent way which currently cannot be corrected due to insufficient knowledge about the value of $D_T$. Choosing to calibrate with the mean would introduce additional bias due to the presence of delta rays, but would make the calibration insensitive to diffusion. To state it succinctly, we can either choose to have a large flat bias, or a smaller but drift-dependent bias. To minimise all bias one could  use the mean value for the response-flattening part of the calibration, and use the MPV for muons close to the anode for the energy scale calibration. In this way we would only introduce a flat bias into response flattening, and this would be corrected when setting the energy scale, with some small residual uncertainty from the effects of diffusion on the d$Q$/d$x$ distributions close to the anode. We emphasise, however, that this avenue should be explored in detector-dependent contexts, as reconstruction pathologies, which may also affect the shape of the distribution, may differ for different experimental setups. If the mean cannot be used because reconstruction pathologies are too severe, then one could use the simulation to estimate the true values of the median and MPV as a function of drift time, though this would require simulating $D_T$ accurately and with precision in a way that has not been achieved so far in LArTPCs (appendix B, reference \cite{MicroBooNE:2021icu}).

\subsection{Impact of Diffusion on Particle Identification}

To quantify how the changing shape of the d$E$/d$x$ distribution impacts the particle identification capabilities of LArTPCs, one must choose a particle identification method. Here we choose the $\chi^2$ method, in which one calculates

\begin{equation}
    \chi^2_s = \frac{1}{NDF}\sum^{\mathrm{N. depositions}}{\frac{(\frac{\mathrm{d}E}{\mathrm{d}x}^{obs.}(DTE)-\frac{\mathrm{d}E}{\mathrm{d}x}^{exp.}_{s}(DTE))^2}{\sigma^2}},
\end{equation}

\noindent where $\frac{\mathrm{d}E}{\mathrm{d}x}^{obs.}(DTE)$ is the observed energy deposition per unit length at a given distance to the track end (DTE), and $\frac{\mathrm{d}E}{\mathrm{d}x}^{exp.}_{s}(DTE)$ is the expected energy deposition per unit length at that DTE for a given particle species, $s$. This technique is both simple and ubiquitous in the LArTPC community \cite{ArgoNeuT:2018und,DUNE:2020cqd}, and so is chosen as our baseline. We choose $\sigma^2=\frac{\mathrm{d}E}{\mathrm{d}x}^{obs.}(DTE)$, which is expected to be an overestimate of the uncertainty. 

\begin{figure}
    \centering
    \includegraphics[width=1.0\textwidth]{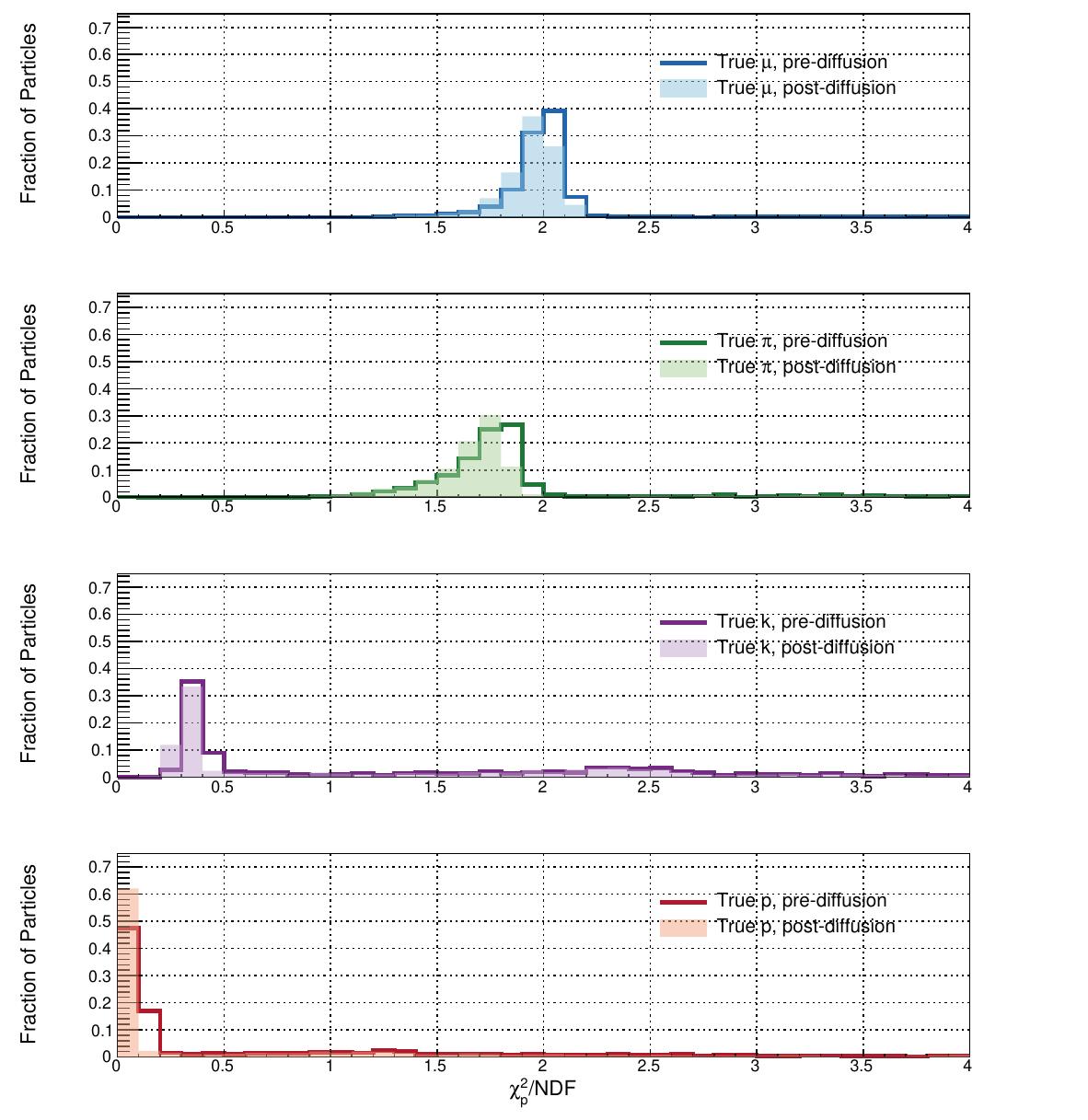}
    \caption{Distributions of $\chi^2_p$ for muons (top), pions (second), kaons (third) and protons (bottom) using the last 30 cm of each track before and after simulating diffusion.}
    \label{fig:chi2p_splitout}
\end{figure}

\begin{figure}
    \centering
    \includegraphics[width=0.6\linewidth]{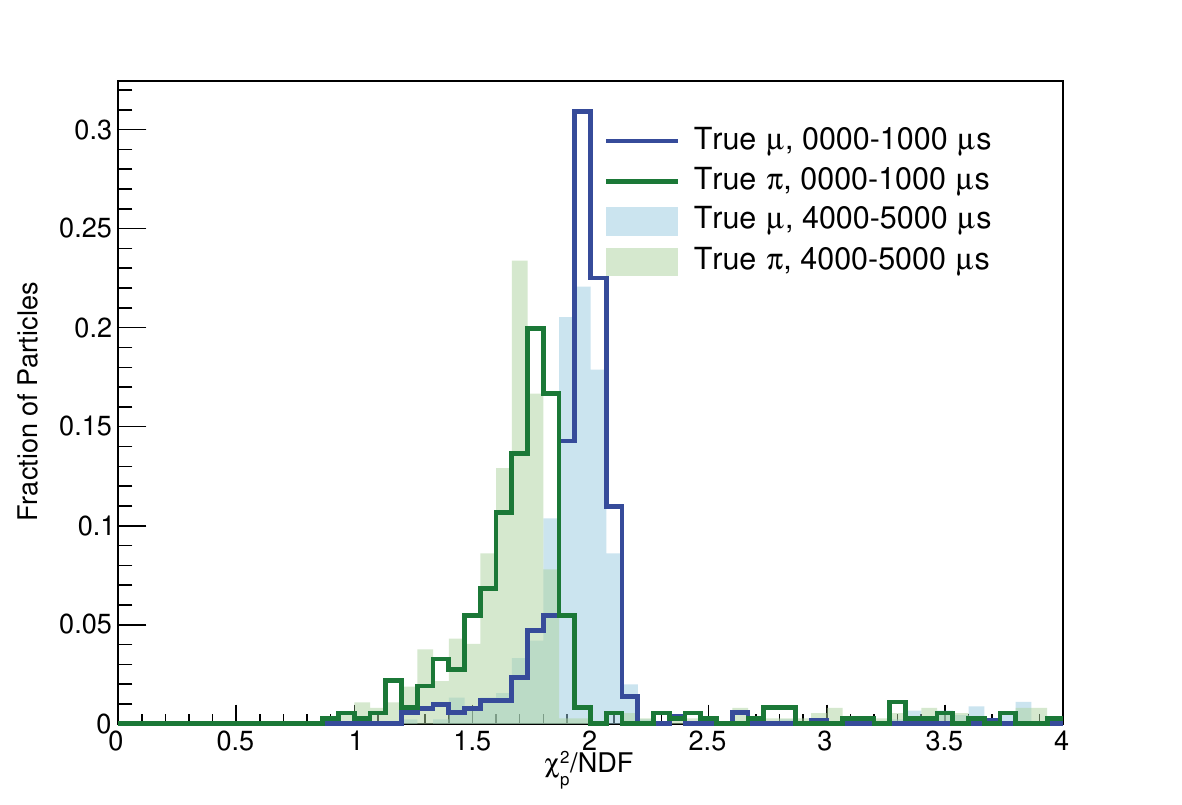}
    \caption{Distributions of $\chi^2_p$ for muons (blue) and pions (green) with diffusion simulated. The line histograms show the distribution for particles at low drift-times while the filled histograms show the distribution for those particles at high drift-times. There is a slight shift to lower PID scores as a function of drift time, indicating a drift-time dependent PID may help to distinguish the two populations.}
    \label{fig:chi2p_pivmu}
\end{figure}

Using the physics-on sample outlined in \S\ref{subsec:samples} we produce distributions of $\chi^2_p$ for muons, pions, kaons and protons using the last 30 cm of the tracks (figure \ref{fig:chi2p_splitout}). The long tails present in the distributions of pions, kaons, and protons are due to events where the particles re-interact before stopping, meaning there is no Bragg peak present and making this PID implementation ineffective. There is a shift to lower scores for all distributions after simulating diffusion, which can be understood as being due to the averaging effect of diffusion on the distributions of $\mathrm{d}E/\mathrm{d}x$. Effectively, the MPV of the distribution moving towards the mean of the distribution results in more depositions having a lower $\chi^2$ when compared to the proton Bragg Peak. Diffusion clearly has a limited effect on the ability of LArTPCs to separate muons from protons, as the change in PID score after applying diffusion is not on the scale of the separation of the two populations. For separating muons from pions, or kaons from protons, the change in PID score due to diffusion is on the scale of the separation of the two populations, so must be considered as a potentially significant source of systematic uncertainty. Figure \ref{fig:chi2p_pivmu} shows the distribution of $\chi^2_p$ scores for muons and pions for tracks near to the anode (0-1000 $\mu$s), and those near to the cathode (4000-5000 $\mu$s). The slight shift to lower PID scores for longer drift times indicates that implementing PID in a drift-time dependent way may help to improve separation of muons and pions, though the effect is small, and current detector resolutions may wash out this effect in data.

In addition to the $\chi^2$ method, we investigated the effect of diffusion on $\mu$-p separation using the more sophisticated algorithm presented in \cite{MicroBooNE:2021ddy}, and found that, similarly, values of the PID score changed by small amounts relative to the separation of the populations. 

\section{Conclusions}

In this work, we have studied how diffusion impacts energy reconstruction and calibration, and particle identification, using a GEANT4 simulation of a LArTPC. There are three primary conclusions from this study. 

The first and most important conclusion is that diffusion can meaningfully impact LArTPC calibrations when the median and MPV $\mathrm{d}Q/\mathrm{d}x$ are used to calibrate. The mean d$Q$/d$x$ incurs a large bias from effects such as delta rays, but is independent of the value of the transverse diffusion coefficient, and so is flat in drift time, making it a good candidate to use to flatten detector response. The MPV of the distribution is likely still the best candidate for setting the energy scale, though ideally the d$Q$/d$x$ distribution would be taken from muons travelling close to the anode to minimise the impact of bias from diffusion. Of course, inclusion of additional detector-specific effects ($E$-field non-linearities, wire response, reconstruction) may render all approaches biased, and so we also suggest an alternative approach; one could use the simulation to extract the true post-diffusion distribution median and MPV to correct to. The precision of this strategy is however limited by the uncertainty on $D_T$.

The second take away is that outside of the impact on calibration, diffusion is unlikely to have an impact on the ability of LArTPCs to separate muons and protons. This conclusion is expected given that diffusion has not shown itself to be source of large systematic uncertainty in currently published analyses using data from LArTPCs. We do find, however, that diffusion may be a significant source of systematic uncertainty for separating muons from pions and kaons from protons. In this case, there is some indication from this study that a drift-time dependent PID could be used to improve the separation of these populations, though further study with a full Monte-Carlo simulation and data is needed.

The third take away is that one must be careful when simulating diffusion, especially with respect to grouping ionization electrons into packets. Doing so can significantly modify $\mathrm{d}E/\mathrm{d}x$ distributions in the simulation in a way that is not present in the data.

Finally, we note that the effects we have presented in this document are entirely driven by transverse diffusion, for which there are still no measurements at the $E$-fields relevant for LArTPCs. Measuring $D_T$ should be a priority to understand diffusion for long-drift distance LArTPCs. 

\section{Acknowledgements}
We would like to thank Vincent Basque, Thomas Carroll, Brian Rebel, and Tingjun Yang for providing their thoughts on this manuscript and many useful conversations. 

%%%%%%%%%%%% BIBLIOGRAPHY %%%%%%%%%%%%%%
\newpage
\bibliographystyle{jhep}
\bibliography{main}

\newpage

%%%%%%%%%%% APPENDICES %%%%%%%%%%%%%%%%

\appendix
\section{Calibration Bias With Physics-On Sample}
\label{app:calibrationwithphysicson}

\begin{figure}[ht]
\centering
\includegraphics[width=1.0\textwidth]{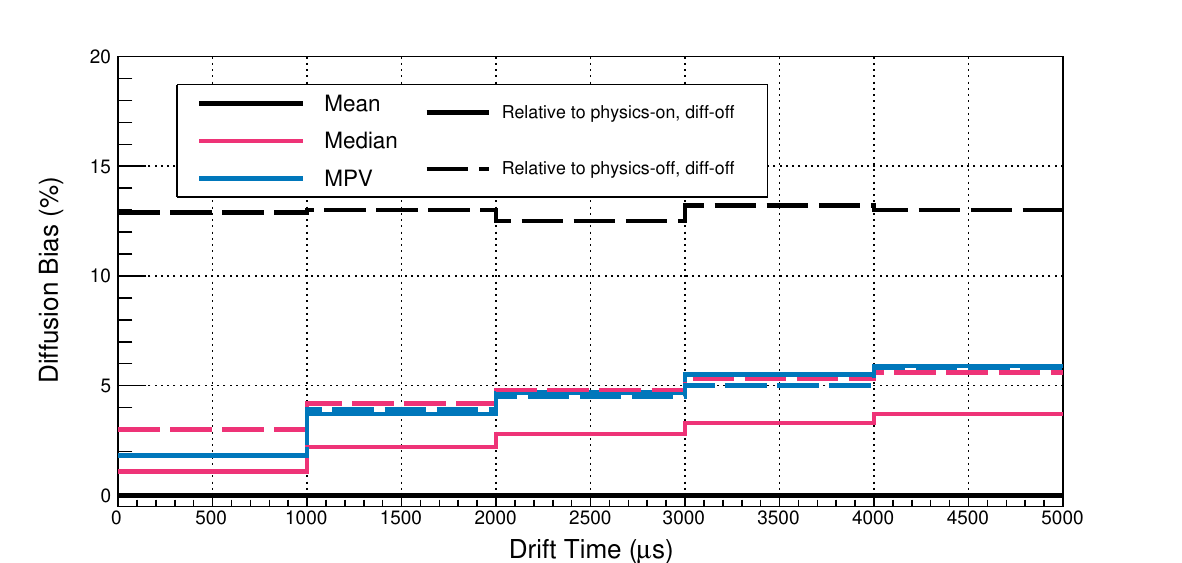}
\caption{Bias of mean, median, and MPV d$Q$/d$x$ values after applying diffusion when compared to the true mean, median and MPV pre-diffusion from the physics-on sample (solid) and physics-off sample (dashed).}
\label{fig:biasplotdeltarays}
\end{figure}

For the calibration studies carried out in this work we used an idealized sample with most physics effects turned off so that we were able to separate out the effects of diffusion from other effects. The study is repeated using the physics-on sample outlined in \S\ref{subsec:samples}, and the results are presented in table \ref{table:meanmeddedxvalsdeltarays}  and shown graphically in figure \ref{fig:biasplotdeltarays}. In general the same conclusions drawn in the main text of this work hold here, though with some notable differences. The bias in the median and MPV of the distributions as a function of drift time are increased relative to the idealized sample, while the mean of the distribution remains unbiased by the effects of diffusion. There is a relatively large bias ($\sim13$\%) incurred when comparing the mean values with physics-on with those values reported in table \ref{table:meanmeddedxvals}, and this difference likely comes from the presence of delta rays, which modify the mean more than either the median or the MPV due to the presence of long tails in the distribution. Notably, this bias is constant across the drift. The fluctuations around the mean bias of $\sim13\%$ are due to comparing statistically independent samples, and similar variations are also present in both the median and MPV curves, though are more difficult to pick out due to the bias changing as a function of the drift time. The average increase in the bias across the drift time from using the physics-on sample are $12.92\pm0.23$ for the mean, $1.96\pm0.05$ for the median, and $0.08\pm0.25$ for the MPV, where the quoted uncertainties are just the RMS of the bin values.

\begin{table}[h]
\begin{center}
\begin{adjustbox}{width={\textwidth},totalheight={\textheight},keepaspectratio}%
\begin{tabular}{c||c|c|c||c|c|c||c|c|c}
      Drift Time ($\mu$s) & \multicolumn{3}{|c||}{Mean d$Q$/d$x$ (e$^-$/ cm)} & \multicolumn{3}{|c||}{Median d$Q$/d$x$ (e$^-$/cm)} & \multicolumn{3}{|c}{MPV d$Q$/d$x$ (e$^-$/cm)}\\
     \hline
     & no diff & diff & \% bias & no diff & diff & \% bias & no diff & diff & \% bias\\
    \hline
    \hline
    0-1000     & 61393.3 & 61385.4 & 0.0 & 52496.3 & 53084.1 & 1.1 & 49046.6 & 49916.7 & 1.8\\
    1000-2000  & 61358.0 & 61348.7 & 0.0 & 52513.7 & 53651.6 & 2.2 & 49030.3 & 50851.2 & 3.7\\
    2000-3000  & 61117.0 & 61125.5 & 0.0 & 52514.6 & 53985.6 & 2.8 & 48856.3 & 51160.6 & 4.7\\
    3000-4000  & 61462.6 & 61462.6 & 0.0 & 52536.0 & 54278.6 & 3.3 & 48742.4 & 51442.0 & 5.5\\
    4000-5000  & 61458.2 & 61399.4 & 0.0 & 52507.9 & 54416.9 & 3.6 & 48800.5 & 51665.5 & 5.9\\
\end{tabular}
\end{adjustbox}
\end{center}
\caption{Mean, median, and MPV d$Q$/d$x$ values before and after applying diffusion for different drift time ranges. These were made using the physics-on sample of muons described in \S\ref{subsec:samples}.}
\label{table:meanmeddedxvalsdeltarays}
\end{table}

\end{document}